\documentclass[a4paper]{jpconf}
\usepackage{graphicx}
\usepackage{amssymb,amsmath,bm,bbm}

\begin{document}
\title{K and B meson mixing in warped extra dimensions with custodial protection}

\author{Bjoern Duling}

\address{Physik Department, Technische Universit\"at M\"unchen, D-85748 Garching, Germany}

\ead{bduling@ph.tum.de}

\begin{abstract}
After a brief theoretical introduction of the warped extra-dimensional model with custodial protection the results  of \cite{Blanke:2008zb} are presented. In this work we analyze the impact of Kaluza-Klein (KK) gauge boson modes on $\Delta F=2$ observables, for the first time considering the full
operator basis and including NLO renormalization group running. It is pointed out that
the dominant contribution in the B-system does not come from the KK
gluon, but that contributions from KK excitations of the weak gauge
bosons are competitive. 
In a numerical analysis we assess the amount of fine-tuning necessary for obtaining realistic values for quark masses and mixings and at the same time realistic values for $\epsilon_K$,
the measure for CP violation in K meson mixing. We are able to show that a mass of the lightest KK gauge boson of 2-3 TeV, and hence in the reach of the LHC, is still possible for moderate fine-tuning. These results enable us to make predictions for not yet measured $\Delta F=2$ observables, such as $S_{\psi\phi}$ and $A_s^{SL}$, which can differ significantly from their SM values.
\end{abstract}

\section{Introduction}
\label{sec:Introduction}
Despite the impressive agreement of the Standard Model (SM) of particle physics with everything we know about the interactions of elementary particle physics, one might be bothered by the large hierarchies that have to be put into the SM by hand.
The large hierarchy between the electro-weak (EW) scale and the Planck scale as well as the large hierarchies in fermion masses and mixing angles cannot be explained within the framework of the SM and hence are referred to as the large hierarchy problem and the flavor problem.

Models with a warped extra dimension, first proposed by Randall and Sundrum (RS) \cite{Randall:1999ee}, in a very ambitious way address both these problems by allowing SM fields, except for the Higgs boson, to propagate into the five-dimensional bulk \cite{Gherghetta:2000qt,Chang:1999nh,Grossman:1999ra}. These models provide a geometrical explanation of the hierarchy between the Planck scale and the EW scale and allow for a natural generation of the hierarchies in the fermion spectrum and mixing angles \cite{Gherghetta:2000qt,Grossman:1999ra}, while simultaneously suppressing flavor changing neutral current (FCNC) interactions \cite{Huber:2003tu,Agashe:2004cp}. Recently realistic RS models of EW symmetry breaking have been constructed \cite{Agashe:2003zs,Csaki:2003zu,Agashe:2004rs,Cacciapaglia:2007fw,Contino:2006qr,Carena:2006bn}, and one can even achieve gauge coupling unification \cite{Agashe:2002pr,Agashe:2005vg}. The models that we will analyze in the following are based on an enlarged bulk gauge group given by

\begin{equation}
G_\text{bulk}=SU(3)_c\times SU(2)_L\times SU(2)_R\times U(1)_X\times P_{LR}\,.
\label{eq:gauge-group}
\end{equation}
The fermions in these models are embedded into representations of $G_\text{bulk}$ in such a way that the $T$ parameter \cite{Agashe:2003zs,Csaki:2003zu} and the $Zb_L\bar b_L$ coupling \cite{Agashe:2006at} are protected from too large corrections.
The protection of these EW observables allows for Kaluza-Klein masses of order $M_\text{KK}\simeq(2-3)\,\textrm{TeV}$ which are in the reach of the LHC \cite{Cacciapaglia:2006gp,Contino:2006qr,Carena:2007ua,Djouadi:2006rk,Bouchart:2008vp}.

As has been pointed out in \cite{Blanke:2008zb,Blanke:2008yr}, the above-mentioned custodial protection of $Zb_L\bar b_L$ also suppresses the flavor off-diagonal $Zd_L^i\bar d_L^j$ couplings as well as the flavor off-diagonal couplings of the additional $Z^\prime$ gauge boson.

Since in this class of RS models well measured FCNC processes related to particle-antiparticle mixings in the $K^0-\bar K^0$ and $B_{d,s}^0-\bar B_{d,s}^0$ systems are known to receive dangerous contributions from KK gluon and EW gauge boson exchanges \cite{Agashe:2004cp,Burdman:2003nt}, the question arises whether these models can be made consistent also with $\Delta F=2$ data for low KK masses of order $M_\text{KK}\simeq(2-3)\textrm{TeV}$.

While the full analysis of $\Delta F=2$ processes in the model under consideration can be found in \cite{Blanke:2008zb}, in these proceedings we will merely try to highlight and illustrate the most important results obtained therein. First, that contradicting naive intuition the impact of EW gauge bosons on $\Delta B=2$ observables can well compete with the impact of KK gluons. This is in contrast to the K system, where as expected the by far dominant contribution comes from KK gluons alone. Second, that despite a generic difficulty in obtaining viable values for $\epsilon_K$ in a  model with anarchic Yukawa couplings, as was pointed out in \cite{Csaki:2008zd}, this is nonetheless possible for a sizable share of the parameter space if a moderate fine-tuning is accepted. Third, that the model is fully realistic and in agreement with all available $\Delta F=2$ data, but still allows for large deviations from the SM in $B_{d,s}$ observables. The rest of this work is organized as follows. After a brief description of the model in Section \ref{sec:Model}, we present our analysis of $\Delta F=2$ observables in Section \ref{sec:Analysis}, motivating as many results as possible in a semi-analytic way. In Section \ref{sec:Conclusions} we present our conclusions. 

\section{The Model}
\label{sec:Model}
The model we are considering is based on the Randall-Sundrum geometric background, that is on a 5D spacetime where the extra dimension is compactified to the interval $0\leq y\leq L$, with a warped metric given by \cite{Randall:1999ee}
\begin{equation}
ds^2=e^{-2ky}\eta_{\mu\nu}dx^\mu dx^\nu-dy^2\,.
\label{eq:metric}
\end{equation}
The curvature scale $k$ is assumed to be of order $k\sim \mathcal{O}(M_\text{Pl})$. Since effective energy scales depend on the position $y$ along the extra dimension via the warp factor $e^{-ky}$, setting $e^{kL}=10^{16}$, or equivalently $kL\approx36$, allows to address the gauge hierarchy problem if the Higgs field is localized close to or at the IR brane ($y=L$). The only free parameter coming from space-time geometry then is given by 
\begin{equation}
f=ke^{-kL}\sim\mathcal{O}(TeV)\,,
\label{eq:f}
\end{equation}
the mass scale of the lightest KK excitations present in the model.\\

At first sight, the most important contribution to $\Delta F=2$ processes seems to come from KK gluons, in particular from the first KK excitation. The profile of the KK gluon along the fifth dimension is given by
\begin{equation}
g(y)=\frac{e^{ky}}{N}\left[J_1\left(\frac{M_\text{KK}}{k}e^{ky}\right)+bY_1\left(\frac{M_\text{KK}}{k}e^{ky}\right)\right]\simeq\frac{e^{ky}}{N}J_1\left(\frac{M_\text{KK}}{k}e^{ky}\right)\,,
\label{eq:gluon-shape}
\end{equation}
where $J_1(x)$ and $Y_1(x)$ are the Bessel functions of the first and second kind, $b\simeq0$ is determined by the boundary conditions at $y=0,L$ and $N$ is a normalization constant. The mass $M_\text{KK}$ of the lightest KK gluon can be numerically determined to be \cite{Agashe:2007ki}
\begin{equation}
M_\text{KK}\simeq 2.45f\,,
\end{equation}
with $f$ being defined in (\ref{eq:f}). 

However, as we will see later, also
some of
the EW gauge bosons $Z, Z_H,Z^\prime$ and the KK photon $A^{(1)}$ arising from the enlarged gauge group given in (\ref{eq:gauge-group})
will have an important impact especially on $\Delta F=2$ observables in the $B_{d,s}$ systems.\\

Allowing the fermions to propagate into the bulk allows for a natural explanation of the observed hierarchies in fermion masses and mixings \cite{Gherghetta:2000qt,Grossman:1999ra,Huber:2003tu} and provides at the same time a powerful suppression mechanism for FCNC interactions, the so called \textit{RS-GIM mechanism} \cite{Agashe:2004cp}.

The bulk profile of a fermionic zero mode depends strongly on its bulk mass parameter $c_\Psi$. In case of a left-handed zero mode $\Psi_L^{(0)}$ it is given by \cite{Gherghetta:2000qt,Grossman:1999ra}
\begin{equation}
f_L^{(0)}(y,c_\Psi)=\sqrt{\frac{(1-2c_\Psi)kL}{e^{(1-2c_\Psi)kL}-1}}e^{-c_\Psi ky}
\label{eq:fermion-shape}
\end{equation}
with respect to the warped metric. For $c_\Psi>1/2$ ($c_\Psi<1/2$) the fermion is localized towards the UV (IR) brane and exponentially suppressed on the IR (UV) brane. Hence a slight hierarchy in the bulk mass parameters of the different quark flavors will lead to a large hierarchy in their respective overlaps with the IR-localized Higgs field and hence to naturally large hierarchies in masses and mixing angles. The details of this construction can be found e.g. in \cite{Albrecht:2009}. It is however important to note at this point that the shape functions for quarks (\ref{eq:fermion-shape}) and KK gauge bosons (\ref{eq:gluon-shape}) will lead to FCNC already at the tree level. The coupling of a zero mode quark $\Psi_{L,R}^{(0)}$ to e.g. the KK gluon in the flavor eigenbasis is given by
\begin{equation}
-i\gamma^\mu t^a\frac{g_s^{5D}}{L^{3/2}}\int\limits_0^L\!dy\ e^{ky}\left[f_{L,R}^{(0)}(y,c_\Psi)\right]^2g(y)\,,
\label{eq:gluon-coupling}
\end{equation}
with $g_s^{5D}$ the 5D $SU(3)_c$ gauge coupling constant. Note that the flavor universality is strongly violated due to the dependence of the overlap integral on the bulk mass parameter $c_\Psi$. When going to the mass eigenbasis via unitary rotations of the quark fields, the non-universality will give rise to FCNCs at the tree level mediated by KK gluons and all other KK gauge bosons, as advertised above. Given this fact it is obvious that the model at hand does not fall into the class of minimal flavor violating (MFV) models \cite{D'Ambrosio:2002ex,Buras:2000dm,Buras:2003jf}, and thus in principle significant deviations from the SM are possible.

\section{Analysis of $\Delta F=2$ observables and CP-violation}\label{sec:Analysis}
\subsection{Theoretical framework and semi-analytic results}
\begin{figure}
\begin{center}
\includegraphics[width=3in]{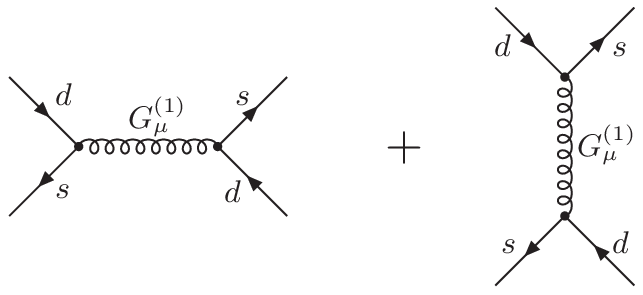}
\end{center}
\caption{\label{fig:tree-level}Tree level contributions of KK gluons to $K^0-\bar K^0$ mixing.}
\end{figure}

We will begin our discussion of $\Delta F=2$ processes with the tree level exchanges of the lightest KK gluons as shown in the case of $\Delta S=2$ transitions in Figure~\ref{fig:tree-level}. Analogous diagrams contribute to $B_{d,s}^0-\bar B_{d,s}^0$ mixings. Tree level EW contributions will be analyzed subsequently.

In addition to the $Q_1^{VLL}=(\bar sd)_{V-A}(\bar sd)_{V-A}=\left[\bar s\gamma_\mu P_Ld\right]\otimes\left[\bar s\gamma^\mu P_Ld\right]$ operator present in the SM, the tree level exchanges of KK gluons give rise to three additional operators: 
\begin{eqnarray}
Q_1^{VRR}&=&(\bar s\gamma_\mu P_Rd)(\bar s\gamma^\mu P_Rd)\nonumber\\
Q_1^{LR}&=&(\bar s\gamma_\mu P_Ld)(\bar s\gamma^\mu P_Rd)\nonumber\\
Q_2^{LR}&=&(\bar s P_Ld)(\bar s P_Rd)\,.
\label{eq:new-operators}
\end{eqnarray}
For the $B_{d,s}$ systems, $\bar s$ and $d$ have to be replaced accordingly. Note that the scalar LR operator in the third line of (\ref{eq:new-operators}) at the scale $\mu_s\sim M_\text{KK}$ only arises from the color structure in the case of an exchanged KK gluon, but will not be generated by the exchange of EW gauge bosons. 

The Wilson coefficients of these operators enter the KK contribution $\left(M_{12}\right)_{KK}$ to $M_{12}^K$, the fundamental theoretical quantity in $K^0-\bar K^0$ mixing, in the following way \cite{Blanke:2008zb}:
\begin{eqnarray}
\left(M_{12}^K\right)_{KK}=\frac{1}{12M_\text{KK}^2}m_KF_K^2\cdot\left[\left(C_1^{VLL}(\mu_L)+C_1^{VRR}(\mu_L)\right)B_1^K\right.\nonumber\\
-\left.\frac{1}{2}R^K(\mu_L)C_1^{LR}(\mu_L)B_5^K+\frac{3}{4}R^K(\mu_L)C_2^{LR}(\mu_L)B_4^K\right]^\ast\,,
\label{eq:M12KK}
\end{eqnarray}
where
\begin{equation}
R^K(\mu)=\left(\frac{m_K}{m_s(\mu)+m_d(\mu)}\right)
\label{eq:chiral-enhancement}
\end{equation}
is a chiral enhancement factor that enters when the matrix elements $\left\langle\bar K^0\left|Q_i^{LR}(\mu)\right| K^0\right\rangle$ are taken. $B_i$ are scale dependent $\mathcal{O}(1)$ hadronic parameters known from lattice calculations \cite{Babich:2006bh,Becirevic:2001xt}. The relevant energy scale for $K^0-\bar K^0$ mixing is given by $\mu_L\sim2\textrm{GeV}$. Expressions for $\left(M_{12}^d\right)_{KK}$ and $\left(M_{12}^s\right)_{KK}$ can be obtainend by replacing the relevant indices in (\ref{eq:M12KK}) and (\ref{eq:chiral-enhancement}) (cf. \cite{Blanke:2008zb}), where now the relevant energy scale is given by $\mu_b\sim m_B$.

Numerically, one finds that in the K system the chiral enhancement is very strong, $R^K(\mu_L)\simeq 20$, while on the other hand in the $B_{d,s}$ systems we have $R^{d,s}(\mu_b)\simeq 1$.  Another difference between $K$ and $B_{d,s}$ systems is given by the renormalization group (RG) evolution of $Q_2^{LR}$, for which the anomalous dimension matrices and QCD factors have been calculated in \cite{Ciuchini:1997bw,Buras:2000if} and \cite{Buras:2001ra}. Besides mixing it with $Q_1^{LR}$, RG running significantly enhances this operator. This enhancement is stronger in the $K$ system, since here we evolve from the KK scale $M_\text{KK}$, where the operators are generated, down to the physically relevant scale $\mu_L\sim 2\textrm{GeV}$, whereas in the $B_{d,s}$ systems we only evolve down to $\mu_b\sim4.6\textrm{GeV}$ and the RG effects are particularly strong at lower energies. A detailed analysis of the relative contributions of all four operators, as done in \cite{Blanke:2008zb}, finally shows the following:

\begin{itemize}
 \item In the $K$ system the scalar LR operator $Q_2^{LR}$ is dominant,
 \item in the $B_{d,s}$ systems no single operator dominates and $Q_1^{VLL}$ can well compete with $Q_2^{LR}$.
\end{itemize}

\begin{figure}[htbp]
\begin{center}
\includegraphics[width=3in]{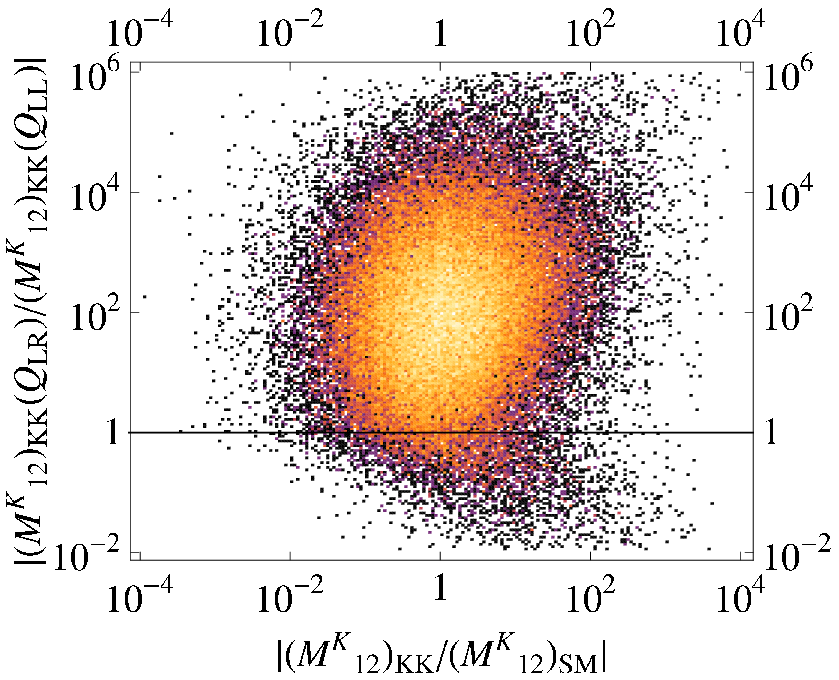}
\includegraphics[width=3in]{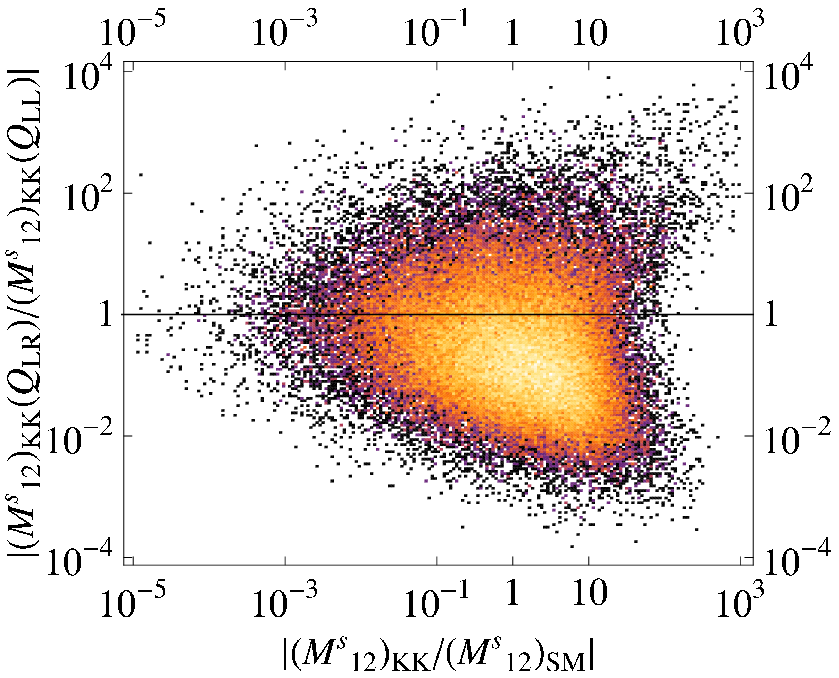}
\end{center}
\caption{\label{fig:M12-LR-LL}The ratio of the contribution of only $Q_{LR}$ and only $Q_{LL}$ to $\left(M_{12}^K\right)_\text{KK}$ (left) and $\left(M_{12}^s\right)_\text{KK}$ (right) as a function of $\left(M_{12}^i\right)_\text{KK}/\left(M_{12}^i\right)_\text{SM}$ ($i=K,s$).}
\end{figure}

This fact is also illustrated in Figure~\ref{fig:M12-LR-LL}, where in the left panel, showing the K system, the ratio of the contribution of only $Q_{LR}$ and only $Q_{LL}$ to $\left(M_{12}^K\right)_\text{KK}$ on average is found to be of order $\mathcal{O}(10^2)$, while in the right panel, showing the $B_s$ system, the analogous ratio is of order $\mathcal{O}(1)$. The situation in the $B_d$ system is very much similar to the one in the $B_s$ system and is not shown here.

The knowledge of the relative importance of the involved operators now allows to estimate the impact of tree level exchanges of EW gauge bosons on $\Delta F=2$ observables in the $K$ and $B_{d,s}$ systems. Just like the KK gluon, also $Z_H$ and $Z^\prime$, that are mainly linear combinations of KK states, have flavor-violating couplings\footnote{Of course this applies also to the KK photon $A^{(1)}$, but its tree level contributions are additionally suppressed by the square of the relevant electric charges and hence negligible.} \cite{Albrecht:2009}. But also the $Z$ boson that receives a small admixture of KK states has flavor-violating couplings, though the coupling to left-handed down-type quarks is suppressed by the custodial protection present in the model\footnote{Another effect that leads to flavor-violating couplings of the Z boson is the mixing of SM quarks with KK quarks, but it is numerically found to be small.} \cite{Blanke:2008zb,Blanke:2008yr}. At first sight, the gluon contribution seems to be by far dominant, since the strong coupling at low energies is much bigger than the EW coupling. But if they are evolved up to energy scales of $\mathcal{O}(M_\text{KK})$, where the additional operators present in the model are generated, these coupling constants become comparable in size and EW contributions could in principle become relevant.
The contributions of tree level exchanges of EW gauge bosons to the various operators relative to the gluon contribution are found to be 

\begin{eqnarray}
C_1^{VLL}(M_\text{KK})&=&\left(1+0.03+0.84\right)\left(C_1^{VLL}(M_\text{KK})\right)_\text{gluon}\nonumber\\
C_1^{VRR}(M_\text{KK})&=&\left(1+0.03+1.46\right)\left(C_1^{VRR}(M_\text{KK})\right)_\text{gluon}\nonumber\\
C_1^{LR}(M_\text{KK})&=&\left(1-0.06-1.69\right)\left(C_1^{LR}(M_\text{KK})\right)_\text{gluon}\nonumber\\
C_2^{LR}(M_\text{KK})&=&\left(C_2^{LR}(M_\text{KK})\right)_\text{gluon}\,,
\label{eq:relative-contributions}
\end{eqnarray}
where the second number in each line refers to the KK photon contribution and the third number to the EW contribution.
Since the dominant operator in $K^0-\bar K^0$ mixing, $Q_2^{LR}$, receives no EW corrections, these are essentially irrelevant in the $K$ system. In the $B_{d,s}$ systems however, where also operators that receive EW corrections are important, the situation is altogether different. Numerically, here the impact of EW gauge bosons can compete with the impact of the KK gluon. This interesting result has been overlooked in the literature and has been first pointed out in \cite{Blanke:2008zb}.\\

\subsection{$\Delta F=2$ analysis and CP-violation in $K^0-\bar K^0$ mixing}
For the following numerical analysis we chose a KK mass scale $M_\text{KK}=2.45\,\textrm{TeV}$ and limited ourselves to those points in parameter space that reproduce the quark masses and the CKM mixing angles to 2$\sigma$, as well as the CKM phase $\gamma$, that still suffers from large uncertainties, to $\pm20^\circ$. To be able to investigate the tension between compatibility with experiment, in particular for $\epsilon_K$, anarchic 5D Yukawa couplings, and a KK mass scale in the reach of LHC, we will in the following assess the amount of fine-tuning present in the model. For this purpose, we choose the fine-tuning definition given by \cite{Barbieri:1987fn},
\begin{equation}
\Delta_{BG}(\mathcal{O})\equiv\max\limits_i\frac{d \log\mathcal{O}}{d \log  x_i}=\max\limits_i\frac{x_i}{\mathcal{O}}\frac{d \mathcal{O}}{d x_i}\,,
\end{equation}
that measures the sensitivity of a given observable $\mathcal{O}$ with respect to the model parameters $x_i$ at a given point in parameter space.

\begin{figure}[htbp]
\begin{center}
\includegraphics[width=3in]{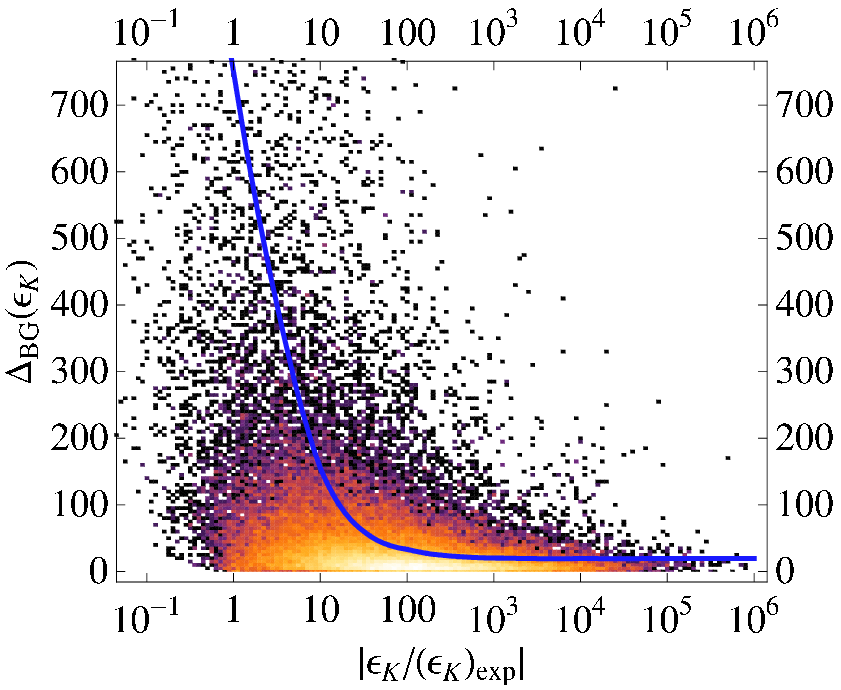}
\includegraphics[width=3in]{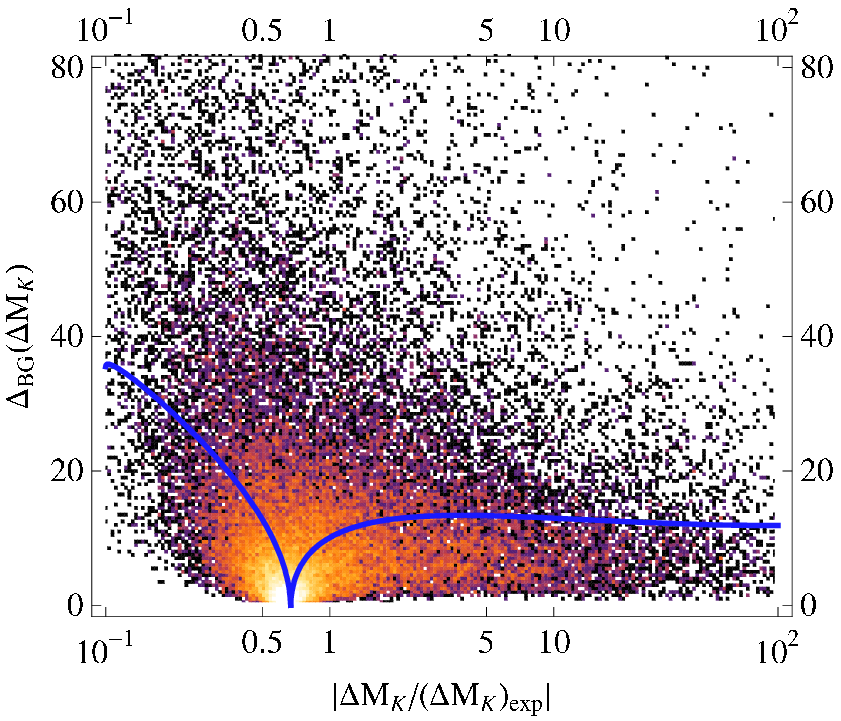}
\end{center}
\caption{\label{fig:epsK-and-DMK}Left: The fine-tuning $\Delta_{BG}(\epsilon_K)$ plotted against $\epsilon_K$, normalized to its experimental value. The blue line shows the average fine-tuning. Right: The fine-tuning $\Delta_{BG}(\Delta M_K)$ plotted against $\Delta M_K$, normalized to its experimental value. The blue line shows the average fine-tuning.}
\end{figure}

The numerical results of our analysis of fine-tuning in $\epsilon_K$ are shown in the left panel of Figure~\ref{fig:epsK-and-DMK}. We can immediately see that generic values for $\epsilon_K$ exceed the experimental value by two orders of magnitude. The reason for this behavior
can be seen in the following way:
The RS-GIM mechanism ensures that for KK scales as low as $M_\text{KK}=2.45\textrm{TeV}$ the typical absolute size of the KK contribution to $M_{12}^K$ does not exceed the size of the SM contribution. In fact, for the chosen KK scale the absolute sizes of these two quantities are roughly equal.
However, since in the SM the imaginary part of $M_{12}^K$, that is responsible for $\epsilon_K$, is accidentally suppressed by two orders of magnitude, the KK contribution to $\epsilon_K$ generically exceeds the SM contribution by the same amount. Deviations of $\epsilon_K$ from this generic value indicate accidental cancellations taking place. Since these accidental cancellations are supposed to be very sensitive to small changes in the model parameters, the fine-tuning in $\epsilon_K$ accordingly increases if $\epsilon_K$ is decreased, such that points in parameter space that satisfy the $\epsilon_K$ constraint display an average fine-tuning in $\epsilon_K$ of about 700.

Having at hand these data, we can determine the KK mass scale that is necessary to suppress the average fine-tuning for $\epsilon\simeq\epsilon_K^{exp}$ below a given value. For a threshold of acceptable average fine-tuning between 10\ -\ 20 we arrive at a generic bound on $M_\text{KK}$ of $15\leq M_\text{KK}\leq25$, thus confirming the result of \cite{Csaki:2008zd}.

Nonetheless, also for $M_\text{KK}=2.45\textrm{TeV}$ we find sizeable areas in parameter space that allow for $\epsilon_K\sim\epsilon_K^{exp}$ with small or moderate fine-tuning, as can be seen in the left panel of Figure~\ref{fig:epsK-and-DMK}. This is due to the fact that the RS model in question has many flavor parameters, and that no less than 14 parameters in the down-type Yukawa sector are involved. Being able to show that the RS model in question despite some inherent tension in $\epsilon_K$ is able to satisfy all experimental constraints from $\Delta F=2$ observables without too much fine-tuning is the central result of \cite{Blanke:2008zb}.\\

For comparison, in the right panel of Figure~\ref{fig:epsK-and-DMK} we also show the fine-tuning in $\Delta M_K$. Here we see that typical values for $\Delta M_K$ are very close to the experimental value, and that the fine-tuning required for obtaining agreement with experimental data is accordingly small. In contrast to $\epsilon_K$, the fine-tuning in $\Delta M_K$ is moderate for all values of $\Delta M_K$. This is even more so the case for all observables in the $B_{d,s}$ systems, showing that the RS-GIM mechanism is very effective for all observables, with the single exception of $\epsilon_K$.\\

\subsection{$\Delta F=2$ analysis and CP-violation in $B_s-\bar B_s$ mixing}

Having at hand results for the most relevant $\Delta F=2$ observables,
and having convinced ourselves that in principle it is possible to obtain agreement with the available $\Delta F=2$ data, we are ready to perform a simultaneous analysis of all available constraints. To this end we impose all $\Delta F=2$ constraints on the RS parameter space. The points we show in the following are all consistent with the experimental data available for $\epsilon_K$, $\Delta M_K$, $S_{\psi K_S}$, $\Delta M_{d,s}$ and $\Delta M_d/\Delta M_s$ and thus fully realistic. In order to maintain naturalness of the theory, we additionally require that the fine-tuning $\Delta_{BG}$ in no observable exceeds 20, still setting $M_\text{KK}=2.45\textrm{TeV}$.

\begin{figure}
\begin{center}
\includegraphics[width=3in]{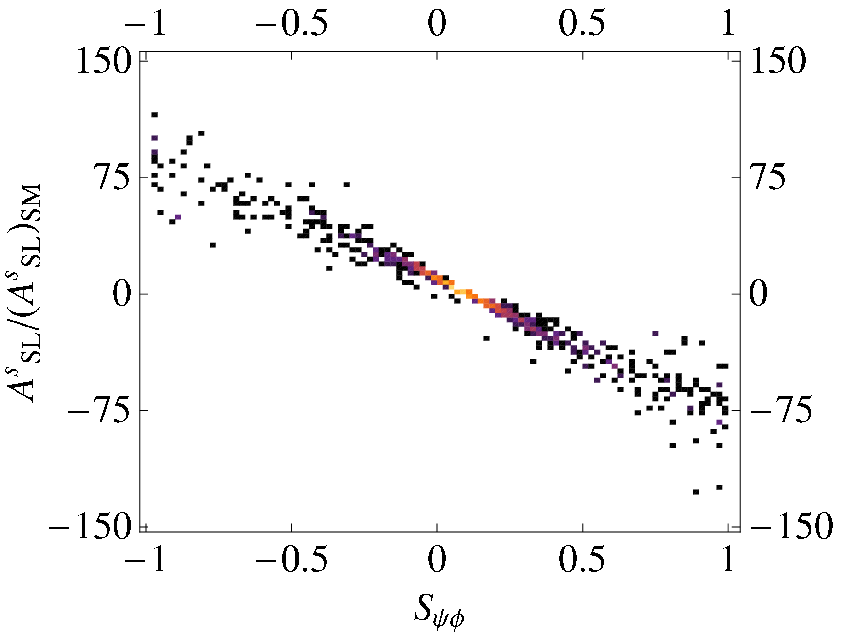}
\includegraphics[width=3in]{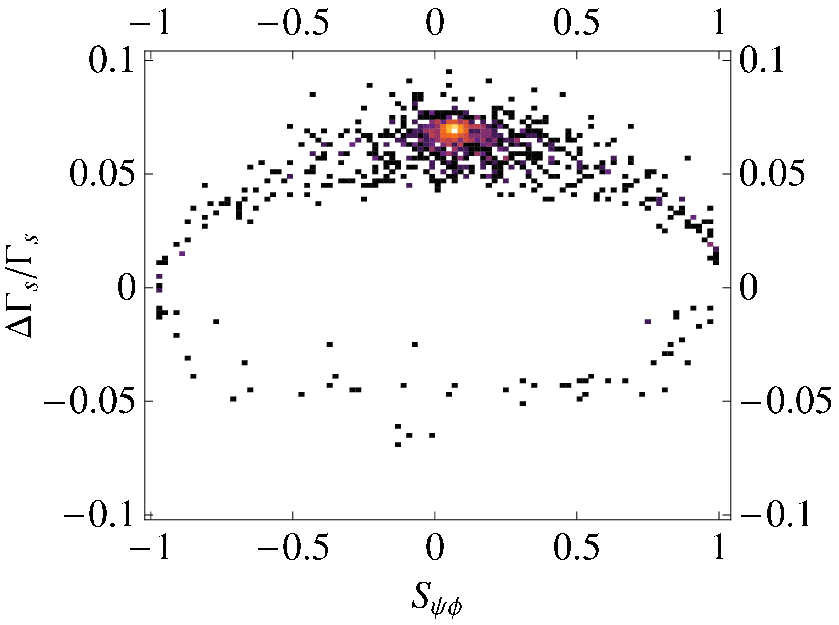}
\end{center}
\caption{\label{fig:AsSL-and-DGs}Left: $A_{SL}^s$, normalized to its SM value, plotted against $S_{\psi\phi}$. Right: $\Delta\Gamma_s/\Gamma_s$, plotted against $S_{\psi\phi}$.}
\end{figure}

In the left panel of Figure~\ref{fig:AsSL-and-DGs} we show the semileptonic CP-asymmetry $A_{SL}^s$ as a function of $S_{\psi\phi}$. We observe that while values of these asymmetries close to the SM ones turn out to be most likely, being a consequence of the RS-GIM mechanism, we find that the full range of new physics phases $\varphi_{B_s}$ is possible, so that $-1<S_{\psi\phi}<1$ to be compared to the SM value $\left(S_{\psi\phi}\right)_\text{SM}\sim0.04$, and also $A_{SL}^s$ can be enhanced by more that two orders of magnitude compared to its SM value. In addition we observe that the model-independent correlation pointed out in \cite{Ligeti:2006pm} turns out to be valid as well in the RS model in question. At this point we would like to mention briefly that the imposition of the fine-tuning constraint, $\Delta_{BG}\leq20$, does not have a qualitative impact on the results stated in this section.

Finally in the right panel of Figure~\ref{fig:AsSL-and-DGs} we show the width difference $\Delta\Gamma_s/\Gamma_s$ as a function of $S_{\psi\phi}$. We observe that due to the correlation between these two observables a future more accurate measurement of $\Delta\Gamma_s/\Gamma_s$ could help to exclude large values of $S_{\psi\phi}$.

\section{Conclusions}
\label{sec:Conclusions}
In \cite{Blanke:2008zb} for the first time the full renormalization group analysis at the NLO level of the most interesting $\Delta F=2$ observables in the $SU(3)_c\times SU(2)_L\times SU(2)_R\times U(1)_X\times P_{LR}$ model has been performed, taking into account the full operator basis and considering simultaneously EW gauge boson and KK gluon contributions. While the presence of the custodial and $P_{LR}$ symmetries in this model ensures consistence with EW precision tests for KK scales as low as $M_\text{KK}\simeq(2-3)\textrm{TeV}$, a recent analysis \cite{Csaki:2008zd} (cf. \cite{Agashe:2008uz,Bauer:2008xb}) points out that for anarchic Yukawa couplings the $\epsilon_K$ constraint requires a much larger KK scale, $M_\text{KK}\simeq(10-20)\textrm{TeV}$. Our detailed analysis confirms these findings; however, having at hand more accurate formulae allows for a quantitative estimate of the fine-tuning that is necessary to reproduce the quark masses and CKM parameters and simultaneously obtain consistency with $\epsilon_K$ and other $\Delta F=2$ observables for lower values of $M_{KK}$. In summary the main results of our analysis are as follows:
\begin{itemize}
 \item While generally $\epsilon_K$ values turn out to be significantly larger than its experimental value, we find regions in parameter space in which the experimental value of $\epsilon_K$ can be reproduced with only moderate fine-tuning.
 \item Interestingly, the tree level EW contributions to $\Delta F=2$ observables, mediated by the additional $Z_H$ and $Z^\prime$ gauge bosons, turn out to be roughly the same size as the KK gluon contributions in the $B_{d,s}$ systems. This is clearly not the case in the K system, where the KK gluon contribution dominates. The $Z$ contributions are of 
higher order in $v^2/M_\text{KK}^2$ 
in both cases and moreover suppressed by the custodial protection of $Zd_L^id_L^j$.
 \item The amount of fine-tuning required to satisfy the $\Delta F=2$ constraints in the $B_{d,s}$ systems is considerably smaller than in the case of $\Delta M_K$ or $\epsilon_K$. This is mainly due to the fact that chiral enhancement as well as RG effects that enhance the dangerous $Q_{LR}$ operators are much smaller in the $B_{d,s}$ systems than in the $K$ system.
 \item Finally, the model allows naturally for $S_{\psi\phi}$ as high as 0.4 as hinted at by the most recent CDF and D\O{} data \cite{Aaltonen:2007he,:2008fj,Brooijmans:2008nt} and by an order of magnitude larger than the SM expectation, $S_{\psi\phi}\simeq0.04$. The strong correlation between $S_{\psi\phi}$ and $A_{SL}^s$ shown in Figure~\ref{fig:AsSL-and-DGs} then would imply a significant departure of the latter observable from its tiny SM value.
\end{itemize}

A detailed analysis of rare $K$ and $B$ decays has been presented in \cite{Blanke:2008yr} and also in \cite{Gori:2009}. A similar analysis of flavor observables in a related model can be found in \cite{Casagrande:2008hr}.

\ack I would like to thank my collaborators Monika Blanke, Andrzej Buras, Stefania Gori and Andreas Weiler as well as Michaela Albrecht and Katrin Gemmler for the fruitful collaboration that led to the papers this work is based on. This work was partially supported by GRK 1054 of Deutsche Forschungsgemeinschaft.

\section*{References}

\providecommand{\newblock}{}

\end{document}